\documentclass[cup6a]{cupbook}
\usepackage{graphicx}

\title[Galaxy evolution as a function of environment and luminosity.]
      {Proceedings of the Workshop held in Vulcano (Messina), Italy,
      May 18-22, 2008}

\author{A. Mercurio, C. P. Haines, A. Gargiulo, F. La Barbera,
      P. Merluzzi and G. Busarello\\INAF-Osservatorio astronomico di
      Capodimonte\\I-80131 Napoli}
\date{\today}

\begin{document}

\pagenumbering{roman}
\maketitle
\tableofcontents
\cleardoublepage
\pagenumbering{arabic}

\chapter{Galaxy evolution as a function of environment and luminosity.}

\section{Introduction}
The properties and evolution of galaxies are strongly dependent on
environment (e.g., \cite{bla05a}, \cite{rin05}, \cite{smi05},
\cite{tan05}). In particular, passively evolving spheroids dominate
cluster cores, whereas in field regions galaxies are typically both
star forming and disc-dominated (\cite{bla05a}). 

However, despite much effort it still remains unclear whether these
environmental trends are: (i) the direct result of the initial
conditions in which the galaxy forms; or (ii) produced later by the
direct interact. Lewis et al. (\cite{lew02}) and Gomez et
al. (\cite{gom03}) that mechanisms such as galaxy harassment or
ram-pressure stripping are not important for the evolution of bright
galaxies. Instead the strongest candidates for driving their
transformation are galaxy suffocation and low-velocity encounters,
which are effective in both galaxy groups and cluster infall regions.
It is not clear if and how this scenario extends to fainter
magnitudes, as there has been observed a strong bimodality in the
properties of galaxies about a characteristic stellar mass $\sim$3
10$^{10}$ M$_{\odot}$ (corresponding to $\sim$M$^{*}$+ 1;
\cite{kau03}). This bimodality implies fundamental differences in the
formation and evolution of giant and dwarf galaxies.  To understand
the mechanisms underlying the transformation of faint galaxies, we
address the relation of galaxy properties with local density and
galaxy mass by analysing low redshift galaxies in SDSS-DR4
(\cite{hai06a}) and the core region of the Shapley supercluster
(\cite{mer06}, \cite{hai06b}).

\section{The data}

We use a volume-limited sample of about 28000 galaxies taken from SDSS
DR4 low-redshift catalogue (LRC) taken from the New York University
Value Added Galaxy Catalogue (NYU-VAGC) of Blanton et al. 2005
\cite{bla05b}, with 0.005$<$z$<$0.037, $\sim$ 90\% complete to
M$_r$=-18.0 (see \cite{hai06a} for details).  For the Shapley
supercluster, we use data from the ESO Archive, acquired with the
ESO/MPI 2.2-m telescope at La Silla. We analysed B- and R-band
photometry of eight contiguous fields covering a 2 deg$^2$ region
centred on the Shapley supercluster core (see \cite{mer06} for
details). The galaxy sample is complete to B = 22.5 ($>$M$^*$+6,
$\mathrm{N_{gal}}$ = 16\,588), and R = 22.0 ($>$M$^*$+7,
$\mathrm{N_{gal}}$ = 28\,008).
 
\section{Results}

Analysing SDSS data, we find that the H$_{\alpha}$ equivalent width,
EW(H$_{\alpha}$), distribution is strongly bimodal, allowing galaxies
to be robustly separated into passively evolving and star-forming
populations about a value EW(H$_{\alpha}$) = 2 \AA.  Examining the
fraction of passively evolving galaxies as a function of both
luminosity and local environment, we find that in high-density regions
$\sim$70\% of galaxies are passively evolving independent of
luminosity (see fig. 1 dashed lines). In the rarefied field, where
environmental related processes are unlikely to be effective, the
fraction of passively evolving galaxies is a strong function of
luminosity, dropping from 50\% for M$_r$=-21 to zero by Mr=-18. Indeed
for the lowest luminosity range covered (-18$<$ Mr $<$ -16) none of
the $\sim$600 galaxies in the lowest-density quartile is passively
evolving. The few passively evolving dwarfs in field regions are
strongly clustered around bright ($\sim$ L$^*$) galaxies, and
throughout the SDSS sample we find no passively evolving dwarf
galaxies more than $\sim$2 virial radii from a massive halo, whether
that be a cluster, group or massive galaxy. Our finding that passively
evolving dwarf galaxies are only found within clusters, groups or as
satellites to massive galaxies indicates that internal processes or
merging are not responsible for terminating star formation in these
galaxies. Instead the evolution of dwarf galaxies is primarily driven
by the mass of their host halo, probably through the combined effects
of tidal forces and ram-pressure stripping.

Moreover the fraction of galaxies with the optical signatures of an
active galactic nucleus (AGN) decreases steadily from $\sim$50\% at
Mr$\sim$-21 to $\sim$0 per cent by Mr$\sim$-18 closely mirroring the
luminosity dependence of the passive galaxy fraction in low-density
environments (see fig. 1 continuous lines). This result reflects the
increasing importance of AGN feedback with galaxy mass for their
evolution, such that the star formation histories of massive galaxies
are primarily determined by their past merger history.

\begin{figure*} 
\centerline{{\resizebox{10cm}{!}{\includegraphics{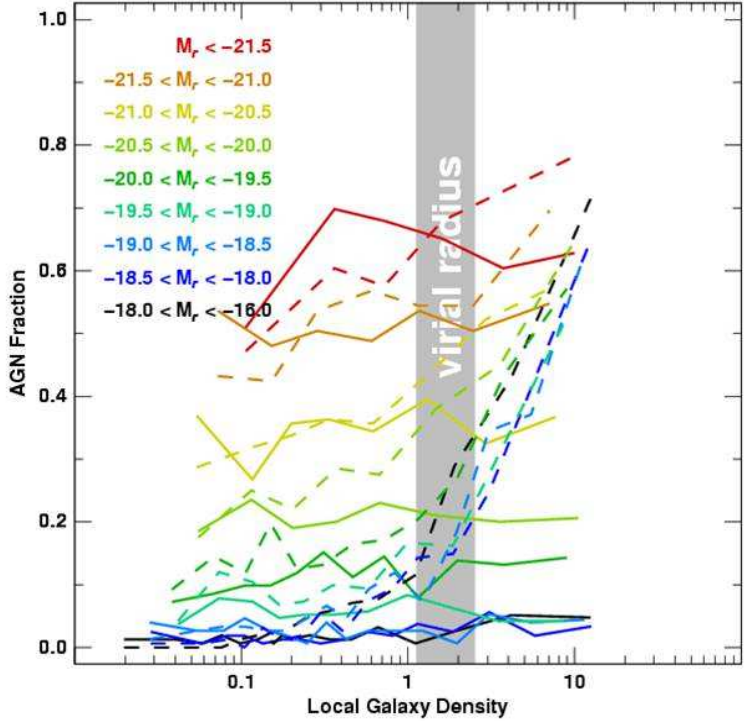}}}}
\caption{The fraction of passively evolving galaxies (dashed lines)
  and of AGN classed from their emission line ratios (continuous
  lines) as a function of both local density and luminosity. Each
  coloured curve corresponds to a different luminosity bin as
  indicated. Each density bin contains 150 galaxies.The grey shaded
  region indicates the typical densities found for galaxies near the
  virial radius in the Millennium simulation.}
\end{figure*}

By analysing optical data of the Shapley supercluster we find that the
galaxy luminosity function cannot be described by a single Schechter
function due to dips apparent at B $\sim$ 17.5 (M$_\mathrm{B} \sim$ -
19.3) and R $\sim$ 17.0 (M$_\mathrm{R} \sim$ - 19.8) and the clear
upturn in the counts for galaxies fainter than B and R $\sim 18$ mag.
We find, instead, that the sum of a Gaussian and a Schechter function,
for bright and faint galaxies respectively, is a suitable
representation of the data. By deriving the galaxy luminosity
functions in three regions selected according to the local galaxy
density, and find that the LF faint-end is different at more than
3$\sigma$ confidence level in regions with different densities.  These
results support the idea that mechanisms related to the cluster
environment, such as galaxy harassment or ram-pressure stripping,
shape the galaxy LFs by terminating star-formation and producing mass
loss in galaxies at $\sim{\mathrm M}^*+2$, a magnitude range where
blue late-type spirals used to dominate cluster populations, but are
now absent. Moreover the observed B-R colour distribution of
supercluster galaxies shows that faint galaxies change from the
cluster cores where $\sim$90\% of galaxies lie along the cluster red
sequence to the virial radius, where the fraction has dropped to just
$\sim$20\%. This suggests that processes directly related to the
supercluster environment are responsible for transforming faint
galaxies, rather than galaxy merging. Their location suggests star
formation triggered by cluster mergers, in particular the merger of
A3562 and the poor cluster SC 1329-313, although they may also
represent recent arrivals in the supercluster core complex.

\begin{figure} 
\centerline{{\resizebox{10cm}{!}{\includegraphics{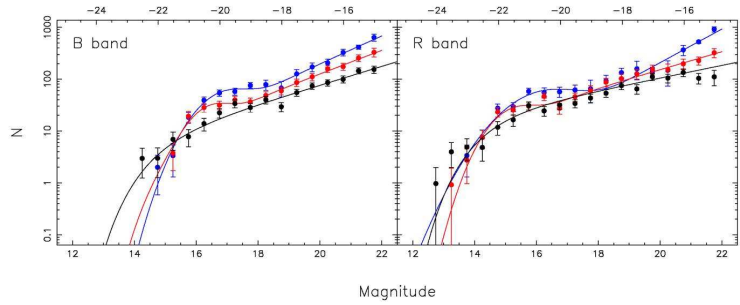}}}}
\caption{The B- (left panel) abd R-band LFs of galaxies in the three
regions corresponding to high- (black), intermediate- (red) and
low-density (blue) environments. Continuous lines represent the best fit
with a Schechter (black) and a Gaussian+Schechter functions.The counts
are  per half magnitude.}
\label{Rlum}
\end{figure}


\end{document}